\documentclass[aps,prb,twocolumn,groupedaddress,floats,showpacs]{revtex4}
\usepackage{latexsym}
\usepackage{dcolumn}
\usepackage[dvips]{graphicx}
\usepackage{amssymb}
\usepackage{graphics}
\usepackage{amsmath}
\usepackage{epsf}

\begin{document}
\author{Shaffique Adam and S. Das Sarma}
\title{Boltzmann transport and residual conductivity in bilayer graphene}
\affiliation{Condensed Matter Theory Center, Department of Physics, 
University of Maryland, College Park, MD 20742-4111, USA}
\date{\today}
\begin{abstract}
A Drude-Boltzmann theory is used to calculate
the transport properties of bilayer graphene.  We find that for 
typical carrier densities accessible in graphene experiments,
the dominant scattering mechanism is overscreened Coulomb impurities
that behave like short-range scatterers.  We anticipate that
the  conductivity $\sigma(n)$ is linear in $n$ at high density and 
has a plateau at low density corresponding to a residual
density of $n^* = \sqrt{n_{\rm imp} {\tilde n}}$, where ${\tilde n}$ 
is a constant which we estimate using a self-consistent
Thomas-Fermi screening 
approximation to be ${\tilde n} \approx 0.01 ~q_{\rm TF}^2 
\approx 140 \times 10^{10}~{\rm cm}^{-2}$.  Analytic results are derived 
for the conductivity as a function of the charged impurity density.
We also comment on the temperature dependence of the bilayer conductivity.    
 \end{abstract}
\pacs{81.05.Uw; 72.10.-d, 73.40.-c}
\maketitle

\section{Introduction}
\label{Sect:Intro}

The recent experimental realization of a single layer of carbon atoms
arranged in a honeycomb lattice has prompted much excitement in both
the theoretical and experimental physics communities (For a recent
review, see Ref.~\onlinecite{kn:dassarma2007a} and references
therein).  The focus of the current work is on bilayer graphene
which has received less attention both theoretically and experimentally,
but is nonetheless of equal importance both for technological
application and for fundamental science.  Bilayer graphene is
two monoatomic layers of graphene separated by about $0.3~{\rm nm}$,
which is the interplane distance in bulk graphite.  Similar to 
single layer graphene, bilayer graphene has been realized 
experimentally through the mechanical exfoliation of graphite 
onto SiO$_2$ substrates.~\cite{kn:dassarma2007a}  While the 
band structure of a single layer of graphene has a linear dispersion, 
theoretically bilayer graphene has a quadratic dispersion 
with an effective mass 
of about $0.03~m_e$ making it similar to the regular
two dimensional electron gas (2DEG).  Despite the quadratic
spectrum, bilayer graphene shares two important features with
single layer graphene (hereafter referred to simply as graphene) 
that distinguish it from regular 2DEGs.  First, the 
bilayer effective Hamiltonian~\cite{kn:mccann2006b,
kn:nilsson2006,kn:partoens2006,kn:koshino2006,kn:snyman2007} is chiral
which gives rise to the anomalous integer quantum hall 
effect.~\cite{kn:novoselov2006}  Second, unbiased bilayer graphene 
is a semimetal implying that one continuously moves
from electron-like carriers for positive gate voltages
to hole-like carriers for negative gate voltages without any
gap in the spectrum.  We note that although recent 
experiments~\cite{kn:oostinga2007} on graphene bilayers have been able to open 
a gap by connecting the upper 
layer to an external top gate, here we ignore this additional 
degree of freedom.~\cite{kn:nilsson2007,kn:min2007}

By considering the gapless bilayer situation, the low density transport
resembles that of single layer where Coulomb impurities
in the substrate create an inhomogeneous density profile breaking
the system into puddles of electrons and holes.  The bulk residual
density $n^*$ induced by these impurities has been calculated for
single layer graphene~\cite{kn:adam2007a,kn:shklovskii2007} which
shows agreement with recent experimental studies.~\cite{kn:tan2007,
kn:chen2007b, kn:martin2007} The high density transport in single
layer graphene with screened Coulomb impurities was discussed in
Refs.~\onlinecite{kn:nomura2007,kn:ando2006,
kn:cheianov2006,kn:hwang2006c,kn:adam2007a,
kn:stauber2007}.  The goal of the present work is to generalize these
high-density and low-density single-layer graphene Boltzmann transport
theories to the case of graphene bilayers.  We note that the Boltzmann
transport theory developed here ignores the effects of phase-coherence
which was studied in
Refs.~\onlinecite{kn:kechedzhi2007,kn:gorbachev2007}.

\section{Bilayer Hamiltonian and Boltzmann transport}
\label{Sect:bihamiltonian}

The effective Hamiltonian for bilayer graphene is now well established
in the theoretical literature (See Refs.~\onlinecite{kn:mccann2006b,
kn:nilsson2006,kn:partoens2006,kn:koshino2006,
kn:snyman2007,kn:katsnelson2006c,kn:katsnelson2007b,kn:cserti2007}).
First principles and band structure calculations show that at both
very small energies and very large energies, bilayer graphene has a
linear spectrum.  For energies $2\times 10^{-3}~{\rm eV} \lesssim
\epsilon \lesssim 0.1~{\rm eV}$ bilayer graphene has a quadratic
spectrum (see e.g. Refs.~\onlinecite{kn:mccann2006b,kn:partoens2006}).  
In principle, bilayer graphene should have a complicated
crossover Hamiltonian moving from linear to quadratic and back 
to linear as
one increased the carrier density.  To our knowledge, a theoretical 
framework to understand the structure and effects of chirality within this
crossover has not yet been developed.  However, as we argue below, charged
impurities in the substrate induce a residual density $n^*$ in 
bilayer graphene that
corresponds to a typical Fermi energy $\epsilon \gtrsim 0.01~{\rm eV}$
which is larger the the lower energy scale for using the quadratic
Hamiltonian, and the range of experimental gate voltages $V_g
\lesssim 50~V$ induces a maximum carrier density with Fermi energy
$\epsilon \lesssim 0.1~{\rm eV}$ which is comparable to the limit
where the low-energy effective quadratic Hamiltonian begins to break down.
Therefore, for realistic samples, the condition $2\times 10^{-3}~{\rm
eV} \lesssim \epsilon \lesssim 0.1~{\rm eV}$ is mostly satisfied and
the quadratic Hamiltonian ${\mathcal H}$ proposed by McCann and Falko
should be an excellent approximation, where in addition bilayer
graphene is weakly interacting in this energy window.  
In what follows, we use~\cite{kn:mccann2006b}

\begin{eqnarray}
{\mathcal H} = -\frac{1}{2 m}\left( \begin{array}{cc} 0 & 
                              \left[p_x - i p_y\right]^2 \\
                              \left[p_x + i p_y\right]^2 & 0
                                   \end{array} \right).
\end{eqnarray}
 
This Hamiltonian can be diagonalized giving $\epsilon_k = \pm \hbar^2
k^2/2m$ where $m = 2 \gamma_1 \hbar^2/(3 \gamma_0^2 a^2) \approx0.033
~m_e$, and $\gamma_0 \approx 3.16~eV$ is the in-plane coupling and
$\gamma_1 \approx 0.39~eV$ is the out of plane coupling, $a \approx
0.246~{\rm nm}$ is the lattice constant and $m_e$ is the electron
mass.  The eigenvectors $\xi_\pm = (e^{-i2\theta_k}, \pm 1)$ give the
aforementioned chiral properties where ${\bf k} = k \exp(i \theta_k)$.

\begin{figure}
\bigskip
\epsfxsize=0.7\hsize
\epsffile{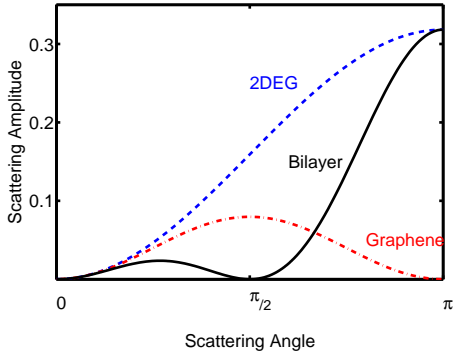}
\caption{\label{Fig:Scat} (Color online) Scattering
cross-section as a function of angle.  Unlike single
layer graphene, both bilayer graphene and 2DEG are dominated
by backscattering.}
\end{figure}

Using this diagonal basis, one can calculate the scattering time
$\tau$ using the Boltzmann transport theory~\cite{kn:ando1982} to find
\begin{equation}
\frac{\hbar}{\tau(\varepsilon_k)} = 2 \pi  \sum_{k'} n_{\rm imp}   
|{\tilde v}(q)|^2 (1 - \hat{{\bf k}}\cdot \hat{{\bf k}}') F_{\hat{{\bf k}}
\hat{{\bf k}'}} \delta(\varepsilon_k - \varepsilon_{k'}),
\end{equation}
where the momentum transfer $q = |{\bf k} - {\bf k}'| = 2 k_{\rm F}
\sin(\theta/2)$, ${\tilde v}(x)$ is the screened scattering impurity
potential and $\hat{{\bf k}}\cdot \hat{{\bf k}}' = \cos \theta$.  The
effects of chirality are captured in $F_{\hat{{\bf k}}\hat{{\bf
k}'}}$.  For a regular 2DEG whose Hamiltonian is not chiral, we have
$F(\theta) = 1$, whereas $F(\theta) = (1 + \cos\theta)/2$ for graphene
and $F(\theta) = (1 + \cos 2\theta)/2$ for bilayer graphene.  The term
$(1 - \hat{{\bf k}}\cdot \hat{{\bf k}}') F_{\hat{{\bf k}}\hat{{\bf
k}'}}$ determines the scattering cross-section which has been plotted
in Fig.~\ref{Fig:Compare}.  We observe that in contrast to single
layer graphene, both bilayer graphene and the 2DEG are dominated by
backscattering ($\theta = \pi$).  Introducing $x = q/(2 k_{\rm F})$,
we find

\begin{eqnarray}
\label{Eq:Boltzmann}
\frac{\hbar}{\tau} = \left\{ \begin{array}{l}
n_{\rm imp} \frac{16 m}{\pi} \left. \int_0^1 dx 
|{\tilde v}(x)|^2 \frac{(x - 2x^3)^2}{\sqrt{1-x^2}} \right.
           \ \ \  \mbox{for bilayers}, \\ \\
n_{\rm imp} \frac{4 \varepsilon_{\rm F}}{\pi v_{\rm F}^2} 
\left. \int_0^1 dx 
|{\tilde v}(x)|^2 x^2 \sqrt{1 - x^2} \right.
           \ \ \  \mbox{for graphene},       
\end{array} \right.
\end{eqnarray}
where the single layer graphene result was reported previously in 
Refs.~\onlinecite{kn:hwang2006c,kn:adam2007a,kn:adam2007b}.  

Before calculating the Boltzmann conductivity $\sigma =  (2 e^2/h)
k_{\rm F} v_{\rm F} \tau$, we first use dimensional arguments to
determine the dependence of conductivity on carrier density.  In 2DEGs
and bilayer graphene the Fermi velocity $v_{\rm F}=\hbar k_{\rm F}/m$
depends on carrier density through $n \sim k_{\rm F}^2$ while the
inverse screening length $q_{\rm TF}=4me^2/(\kappa \hbar^2)$ is
density independent.  Since the gas density parameter $r_s \sim v_{\rm
F}^{-1}$, it scales as $n^{-1/2}$.  This is all in sharp contrast to
single layer graphene where $v_{\rm F} \approx 10^{6}~{\rm m}/{\rm s}$
is constant and $q_{\rm TF} = 4 k_{\rm F} r_s$ depends on carrier
density, and $r_s = e^2/(\kappa v_{\rm F}) \approx 0.8$ is a density
independent constant that depends mostly on the dielectric constant
of the substrate.

One finds that for bilayer graphene $\sigma \sim k_{\rm F}^2 \tau$, and
that for unscreened Coulomb impurities $\tau_C \sim k_{\rm F}^2$
giving $\sigma_C \sim n^2$, whereas overscreened Coulomb scatterers
behave similar to white-noise disorder giving density independent
$\tau$ and $\sigma \sim n/n_{\rm imp}$ (this is similar to a
low-density 2DEG, where Coulomb scatterers are strongly screened if
$q_{\rm TF} \gg 2 k_{\rm F}$).  Writing $\sigma(n) \sim n^{\alpha}$,
we note that $\alpha = 1$ in both the linear and quadratic
Hamiltonians arising from very different reasons (See
Table~\ref{Tab:1}).  The solution within a crossover between a
quadratic and linear Hamiltonian (see above) is beyond the scope of
this work, but we observe that approaching the crossover from either
side gives $\alpha \geq 1$, and for Coulomb scatterers located at the
SiO$_2$ interface, we have $\alpha \leq 2$.  In what follows we focus
on the experimentally relevant regime, where we assume that $q_{\rm
TF}/2 k_{\rm F} >1$ which is typically called the low density regime
in 2DEG literature.~\cite{kn:lilly2003}  In this context, even in GaAs
heterostructures (where $m \approx 0.07~m_e$) moving to higher density
results in a complicated crossover where the exponent $\alpha$ slowly
decreases with increasing density as other scattering mechanisms come
into play.~\cite{kn:lilly2003}

\begin{table}
\caption{\label{Tab:1} Summary of Boltzmann transport
results in 2 $d$ electron gas (2DEG), single layer graphene 
and bilayer graphene.  For screened Coulomb scattering 
results in 2DEG and bilayer graphene we assume 
that $q_{\rm TF}/2k_{\rm F} >1$ (see text), 
and observe that arising from different 
physics, screened Coulomb scattering gives $\sigma \sim n$ 
in all three cases.}
\vspace{0.2in}
\begin{tabular}{||c|c|c|c||}
\hline \hline
& \hspace{0.55in} & \hspace{0.55in}  & \hspace{0.55in}   \\ 
\mbox{} & 2DEG & Graphene & Bilayer \\ 
& & & \\
\hline 
& & & \\ 
Bare Coulomb Scattering 
           & $\sigma \sim n^2$ & $\sigma \sim n$   & $\sigma \sim n^2$ \\ 
& & & \\
\hline
& & & \\
Screened Coulomb
           & $\sigma \sim n$ & $\sigma \sim n$  &  $\sigma \sim n$ \\
& & & \\
\hline
& & & \\
Short-range Scattering 
           & $\sigma \sim n$ & $\sigma \sim \mbox{const}$  
                  & $\sigma \sim n$ \\
& & & \\
\hline
\end{tabular}
\end{table}

In single layer graphene, it was shown by
Refs.~\onlinecite{kn:hwang2006b,kn:barlas2007} that for $q \leq 2 k_{\rm F}$
the static dielectric function calculated in the Random Phase
Approximation is identical to that of the much simpler Thomas-Fermi
approximation.  A similar result holds for 2DEGs.  While it is not
clear if this holds in bilayer where the polarizability
has only been calculated numerically (See
Ref.~\onlinecite{kn:wang2007}), these results indicate that
Thomas-Fermi approximation (which allows for analytical results)
should capture both qualitatively and quantitatively the transport
properties of bilayer graphene.  Within the 
Thomas-Fermi approximation, the potential of a charged impurity
located at a distance $d$ from the substrate is
\begin{eqnarray}
\label{Eq:TF}
{\tilde v}(q) = \frac{2 \pi e^2}{\kappa} \frac{e^{-qd}}{q + q_{\rm TF}} 
\approx \frac{\pi \hbar^2}{2 m},
\end{eqnarray}
where in the second equation we have used the further approximation
(also called ``complete screening approximation'') that $q_{\rm TF} =
4me^2/(\kappa \hbar^2) \sim 1~{\rm nm}^{-1}$ is larger than the
maximum transfered momentum $q \lesssim 0.3~{\rm nm}^{-1}$.  Herein
lies an important difference between single layer graphene and bilayer
graphene.  For single layer graphene $q_{\rm TF} = 4 k_{\rm F} r_s$
depends on density, so that both the screened and unscreened Coulomb
potential scale as $k_{\rm F}^{-1}$.  It is this property of single
layer graphene that gives rise to the conductivity with Coulomb
scatterers being linear in density and the inapplicability of Gaussian
white-noise models (i.e. zero-range scattering) to capture the
transport properties.  In contrast, for bilayer graphene and 2DEG,
$q_{\rm TF}$ is a density independent constant which is larger than
the typical momentum transfered in current experiments, and therefore
the strong screening approximation ($q_{\rm TF} > 2 k_{\rm F}$)
applies except at very high carrier densities.  In this context,
bilayer graphene is much more similar to 2D Si MOSFETs than to single
layer graphene.  Analytic results for the conductivity can be obtained
both in the limit $d\rightarrow 0$ and 
$2 k_{\rm F}/q_{\rm TF} \rightarrow 0$.  Keeping both to leading 
order, we find
\begin{equation}
\sigma(n) \approx \frac{4 e^2}{\pi h} \frac{n}{n_{\rm imp}}
       \left[1 + \frac{1216}{105\sqrt{\pi}} \sqrt{n} (d + q_{\rm TF}^{-1}) \right].
\end{equation}

The result $\sigma = (4 e^2/\pi h) (n/n_{\rm imp})$, where conductivity
is linear in density is valid for $d^{-1}, q_{\rm TF} \gg 2 k_{\rm F}$, 
although
we expect deviations from this linear behavior for $n \geq 6 \times
10^{11}~{\rm cm}^{-2}$, a regime, which in principle should be accessible in
future experiments.  We note that the linear in density behavior 
was anticipated in Ref.~\onlinecite{kn:koshino2006} and in
Ref.~\onlinecite{kn:katsnelson2007b}, but we point out that the low
density saturation in Ref.~\onlinecite{kn:koshino2006} arises from a
completely different and universal
mechanism~\cite{kn:katsnelson2006c,kn:cserti2007} that we believe is
unobservable in current bilayer graphene samples because of the large
and non-universal $n^*$ arising from the disorder induced electron-hole
puddles.~\cite{kn:adam2007a,kn:cheianov2007}  In the following section
we calculate the voltage fluctuations and residual density induced
by charged impurities.  

\section{Low Density Residual Density}

This section follows closely the derivation 
in Refs.~\onlinecite{kn:adam2007a,kn:galitski2007} and applying the
same formalism to 
the case of bilayer graphene.  We consider $N_{\rm imp}$ impurities
located at the points $\{r_{\rm i}\}$ in a 2D plane.  If each impurity 
has a potential $\phi(r)$, then the disorder averaged potential
\begin{eqnarray}
{\bar V} &=& \int d{\bf r}_1 d{\bf r}_2 \cdots 
                P[{\bf r}_1] P[{\bf r}_2] \cdots 
                \sum_{i=1}^{N_{\rm imp}} \phi(r_i), \nonumber \\
         &=& n_{\rm imp} {\tilde \phi}(q=0), 
\end{eqnarray}
where to get the second line we have assumed that the impurities
are uncorrelated and uniformly distributed.  For sample area $A$, 
$n_{\rm imp} = N_{\rm imp}/A$, $P[{\bf r}_i] = A^{-1}$ and 
${\tilde \phi}(q)$ is the 2D Fourier Transform of the impurity 
potential $\phi(r)$.  For example, the real space Coulomb potential
for an impurity located a distance $d$ from the graphene (or bilayer)
plane, $\phi(r) = (e^2/\kappa)[{\bf r}^2 + d^2]^{-1/2}$.  This gives
for the bare potential ${\tilde \phi_o}(q) = (2 \pi e^2/\kappa)\exp(-qd)/q$.
Using the Thomas-Fermi dielectric function $\epsilon(q) = 1 + q_{\rm
TF}/q$ gives the screened Thomas-Fermi potential shown in
Eq.~\ref{Eq:TF}.  Since the static polarizability at $q\rightarrow 0$
is related to the density of states $\nu$ by the compressibility sum
rule, the result for the disorder averaged potential ${\bar V} =
n_{\rm imp}/\nu$ is actually quite general.  For bilayer
graphene $\nu$ is constant, and the threshold voltage shift can be
immediately obtained from ${\bar n} = {\bar V} \nu = n_{\rm imp}$.  It
is the property that $q_{\rm TF}$ is independent of carrier density
that explains why the threshold voltage shift in bilayers is ${\bar n}
= n_{\rm imp}$, which is in contrast to graphene~\cite{kn:adam2007a}
where the density dependent inverse screening length gives ${\bar n} =
n_{\rm imp}^2/4n^*$.  This non-linear dependence of the threshold
voltage on charged impurity density in monolayer graphene has recently
been verified experimentally.~\cite{kn:chen2007b}

The disorder averaged potential fluctuations $\langle V^2 \rangle$ 
are obtained in a similar fashion.
\begin{eqnarray}
\langle V^2 \rangle &=& \int d{\bf r}_1 d{\bf r}_2 \cdots 
                P[{\bf r}_1] P[{\bf r}_2] \cdots 
                \sum_{i,j=1}^{N_{\rm imp}} \phi(r_i) 
                   \phi(r_j), \nonumber \\
\langle V^2 \rangle - {\bar V}^2   &=& \int d{\bf r}_1 d{\bf r}_2 \cdots 
                P[{\bf r}_1] P[{\bf r}_2] \cdots 
           \sum_{i=j=1}^{N_{\rm imp}} \phi(r_i) \phi(r_j), \nonumber \\
        &=& n_{\rm imp} \int \frac{d {\bf q}}{(2 \pi)^2} [{\tilde \phi}(q)]^2.
\end{eqnarray}
For the Thomas-Fermi potential Eq.~\ref{Eq:TF}, we have
\begin{eqnarray}
\langle V^2 \rangle - {\bar V}^2 &=& 
\frac{n_{\rm imp}}{2 \pi} \int q dq 
\left[ \frac{2 \pi e^2}{\kappa}\frac{e^{-qd}}{q + q_{\rm TF}} \right]^2, 
\nonumber \\
&=& 2 \pi n_{\rm imp} (e^2/\kappa)^2 C_0^{\rm TF} (x = 2 q_{\rm TF} d),
\end{eqnarray}
where $C_O^{\rm TF}(x) = \partial_{x} (x e^{x} E_1[x])$ and $E_1[x] =
\int_{x}^{\infty} dt e^{-t}/t$ is the exponential integral function.
While the results presented here are for the Thomas-Fermi potential
Eq.~\ref{Eq:TF}, it is straightforward to generalize this result to
obtain $C_0^{\rm RPA}$ for RPA screening (e.g. using the numerical
bilayer dielectric function calculated in Ref.~\onlinecite{kn:wang2007}), but
for the relevant density scale set by $q_{\rm TF}$, we expect these
results to be quantitatively quite similar. The self-consistent 
density $n^*$ is found by setting $E_{\rm F}^2 =
\langle \delta V^2 \rangle$ (see discussion in
Refs.~\onlinecite{kn:adam2007a,kn:shklovskii2007}) where this
approximation ignores the exchange and correlation contributions which
are believed to be
small.~\cite{kn:hwang2006b,kn:barlas2007,kn:min2007}  Applying
this self-consistent procedure to bilayer graphene, we 
find $(\pi n^*/2m)^2 = 2 \pi n_{\rm imp} (e^2/\kappa)^2 C_O^{\rm TF}$ and 

\begin{eqnarray}
n^* &=& \sqrt{n_{\rm imp} {\tilde n}}, \nonumber \\
{\tilde n} &=& \frac{1}{2 \pi} q_{\rm TF}^2 C_0^{\rm TF}(2 q_{\rm TF} d) \\
 && \approx 140 \times 10^{10} {\rm cm}^{-2} \nonumber,
 \end{eqnarray}
where we approximate $C_O^{\rm TF}(x \approx 1) \approx 0.085$.  This
disorder induced residual density gives rise to a non-vanishing
conductivity even as the external gate voltage is tuned through zero.

\section{Self-consistent Transport Theory}

\begin{figure}
\bigskip
\epsfxsize=0.7\hsize
\epsffile{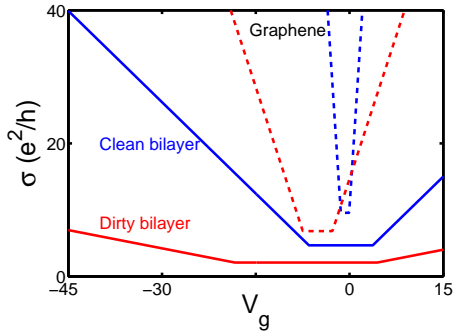}
\caption{\label{Fig:Compare} (Color online) Self-consistent
Boltzmann theory for bilayer graphene (solid lines) compared
with results of Ref.~\onlinecite{kn:adam2007a} for monolayer
graphene (dashed lines).}
\end{figure}

Combining the low and high density results developed in 
the preceding sections, we find for bilayer graphene

\begin{eqnarray}
\label{Eq:Main}
\sigma(n - \bar{n}) = \left\{ \begin{array}{l}
      \frac{4 e^2}{\pi h} \sqrt{\frac{\tilde n}{n_{\rm imp}}}
            \ \ \  \mbox{if $n - \bar{n}< n^*$}, \\
        \frac{4 e^2}{\pi h} \frac{n}{n_{\rm imp}}
             \ \ \ \mbox{if $n - \bar{n} > n^*$},
 \end{array} \right. 
\end{eqnarray}
where $\bar{n} = n_{\rm imp}$ and ${\tilde n} \approx 140 \times
10^{10} {\rm cm}^{-2}$.  This result predicts that a reasonably clean
bilayer sample with $n_{\rm imp} = 5 \times 10^{10} {\rm cm}^{-2}$
would have a mobility $\mu \sim n_{\rm imp}^{-1} \approx 6000~{\rm
cm}^2/Vs$. The residual density $n^* = \sqrt{n_{\rm imp} {\tilde n}}
\sim 25 \times 10^{10} {\rm cm}^{-2}$ with a plateau width $\Delta V
\sim 4 V$ and minimum conductivity $\sigma_{\rm min} \sim 7 e^2/h$.
When compared to recent experimental results,~\cite{kn:morozov2007}
these estimates agree well for the mobility, plateau width and minimum
conductivity, although not for the offset gate voltage determined from
${\bar n}$ (see Refs.~\onlinecite{kn:hwang2006e,kn:schedin2006} for a
discussion of other factors that could determine the threshold voltage
shifts and could account for this discrepancy).  These results do not
depend qualitatively on the precise choice of $d$, although the
results do depend quantitatively; for example, for the same value of
$n_{\rm imp}$, increasing $d$ by a factor of $2$ gives $n^* \approx 15
\times 10^{10} {\rm cm}^{-2}$.  This may be important for bilayer
graphene, since the distance between the two layers $c\sim 0.3~{\rm
nm}$ suggests that for the same substrate, the effective distance from
the charged impurities would be larger for bilayers than for graphene.
The results of Eq.~\ref{Eq:Main} are shown in Fig.~\ref{Fig:Compare}
for both a clean ($n_{\rm imp} = 10^{11} {\rm cm}^{-2}$) and dirty
($n_{\rm imp} = 5 \times 10^{11} {\rm cm}^{-2}$) samples and compared
with the results of Ref.~\onlinecite{kn:adam2007a} for graphene using
the same charged impurity densities and keeping $d = 1~{\rm nm}$
fixed.  One notices immediately that for the same charged impurity
concentration, graphene has a factor of $16$ higher mobility, smaller
plateau widths and larger minimum conductivities than the bilayer
system.  These predictions can be easily tested in future experiments.
We also note that only $\sigma(n) \sim n$ behavior was observed in the
experiments of Ref.~\onlinecite{kn:morozov2007} indicating that the
current experiments are adequately described by the complete screening
approximation, although future experiments should observe a 
``super-linear'' conductivity (i.e. $\alpha > 1$) at higher density 
arising both from the
high-density Thomas-Fermi corrections as well as from the high-density
graphene bilayer Hamiltonian crossing over from quadratic to linear.       

\section{Temperature Dependence}

Shown in the Fig~\ref{Fig:Scat} is the effect of chirality on the
dominant scattering angle, where the suppression of backscattering
seen in graphene is absent for bilayer graphene.  It was argued
recently that the suppression of $2k_{\rm F}$ scattering in graphene
implied weak temperature dependence~\cite{kn:cheianov2006} until
higher temperatures where phonon effects are observed.  The fact that
$2k_{\rm F}$ scattering is not suppressed in bilayer graphene does
not, however, necessarily lead to any screening (or equivalently,
Friedel oscillation) induced strong temperature dependence in the
resistivity.  The temperature dependence in bilayer graphene depends on
three dimensionless parameters: $q_{\rm TF}/2k_{\rm F}$; $T/T_{\rm
F}$; and $T/T_{\rm D}$ where $T_{\rm D} \approx \hbar/2\tau$
is the Dingle temperature and $T_{\rm F}$ is the Fermi temperature.
The temperature dependence from screening will be weak if any one of 
these three parameters is not large.  

The actual value of the dimensionless screening parameter
\begin{eqnarray}
\frac{q_{\rm TF}}{2 k_{\rm F}} && = 
            \frac{2me^2}{\kappa \hbar^2 \sqrt{\pi n}}, \\
&& \lesssim \frac{2me^2}{\kappa \hbar^2 \sqrt{\pi n^*}} \approx 6, \nonumber
\end{eqnarray}
is reasonably small even at the lowest accessible carrier density set
by $n^* \sim 2.5 \times 10^{11}~{\rm cm}^{-2}$, making the temperature
dependence arising from screening rather weak.  Second, the
dimensionless temperature $T/T_{\rm F}$ is rather small since the
Fermi temperature $T_{\rm F}$ changes from $120~K$ at low density (set
by $n^*$) to $1200~K$ at high density (set by $n = k_{\rm F}^2/\pi = a
V_g \approx 3.6 \times 10^{12}~{\rm cm}^{-2}$, where $a \approx 7.2
\times 10^{10}~{\rm cm}^{-2}V$ is a geometry related factor).    

The Fermi temperature is relatively high due to the very small carrier
effective mass ($\approx 0.03~m_e$) in bilayer graphene.  In fact, the
effective mass for bilayer graphene is less than half of that 
for 2D electrons in GaAs  ($\approx 0.07~m_e$), where the temperature 
dependence
arising from screening is extremely small~\cite{kn:lilly2003} even at
much lower carrier densities.  Therefore, we do not anticipate any
strong screening-induced temperature dependence in bilayer graphene
resistivity.  Finally, the currently available bilayer graphene
samples have very small mobilities resulting in relatively strong
collisional broadening effects (i.e. high Dingle temperature) which
would suppress any small screening induced temperature dependence that
could have arisen at low temperatures.  In particular, a mobility
of $5000~{\rm cm}^{2}/Vs$ as observed in Ref.~\onlinecite{kn:morozov2007}     
corresponds to a $T_{\rm D} \sim 50~K$, leading to further suppression
of any screening induced temperature dependence in the conductivity.

It is therefore gratifying to see that the recent experiment on
bilayer graphene~\cite{kn:morozov2007} does not see much temperature
dependence in the low temperature resistivity in spite of the
importance of $2k_{\rm F}$ scattering in bilayer graphene.  The
temperature dependence seen in the plateau region is likely to be
caused by thermal population of carriers since $T_{\rm F} \approx
120~{\rm K}$, and the temperature dependence is seen for $T \gtrsim
100~{\rm K}$.  For $T \gg T_{\rm F}$, the thermally excited 
carrier density 
$n(T) = \int \nu(\varepsilon) f(\varepsilon,T) d\varepsilon $,
where $f(\varepsilon,T)$ is the Fermi distribution function.  This 
gives $n \sim T$ for bilayer graphene, while  $n \sim T^2$ 
for graphene giving for the conductivity (ignoring any phonon
or electron-hole scattering contributions) 
\begin{eqnarray}
\label{Eq:Temp}
\sigma(T\gg T_{\rm F}) = \left\{ \begin{array}{l}
       \frac{8 \ln 2}{\pi^2 h} \frac{me^2}{\hbar^2 n_{\rm imp}} 
           (k T) \ \ \  \mbox{for bilayers}, \\
       \frac{10 \pi e^2}{3 h} 
           \frac{1}{n_{\rm imp} \hbar^2 v_{\rm F}^2} (k T)^2
           \ \ \  \mbox{for graphene}.       
\end{array} \right.
\end{eqnarray}
Note that the thermal excitation of carriers leads
to an enhanced~\cite{kn:morozov2007} $\sigma(T)$  whereas
temperature dependent $2 k_{\rm F}$ screening typically
suppresses $\sigma(T)$.  For bilayer graphene, this result suggests 
that for  $T\approx 260~K$, there would be a $300$ percent enhancement
in the minimum conductivity which is consistent
with the observations and estimates of Ref.~\onlinecite{kn:morozov2007}. The 
situation for single layer graphene is quite different, since
even in the low density saturation regime, $T/T_{\rm F} \ll 1$.  For
example, a residual density $n^* = 2.5 \times 10^{11} {\rm cm}^{-2}$ 
corresponds to  $T_{\rm F} \approx 100~{\rm K}$ for bilayer 
graphene and $T_{\rm F} \approx 700~{\rm K}$ for single layer graphene.
For this reason, one expects the minimum conductivity in bilayer graphene 
to show stronger temperature dependence than single layer graphene even
though Eq.~\ref{Eq:Temp} shows the bilayer conductivity scaling as $\sim T$
(compared to $\sigma \sim T^2$ for the monolayer).

\section{Conclusion}

The formalism developed here captures many features of graphene
bilayer transport.  In particular, we show that both the low density
saturation and the high-density linear in density behavior seen in
recent experiments arise from the same charged impurities that are
invariably present in samples exfoliated onto a SiO$_2$ substrate.  We
note that the agreement between our self-consistent Boltzmann theory
and available graphene (and bilayer) transport experiments and direct
measurement of the electron and hole puddles giving rise to the
residual density, indicate that in current experiments, most of the
charged impurities reside close to the graphene-substrate interface
(i.e. within $\sim 1~{\rm nm}$) similar to the corresponding situation
in Si-SiO$_2$ MOSFET structures.  If the typical distance of
impurities from the graphene (or bilayer) sheet could be changed in
future experiments (e.g. by removing the SiO$_2$ substrate), the
formalism we have developed predicts quantitatively its effect on
transport properties and on the magnitude of the electron and hole
puddles.  Another conclusion of our work is that bilayer mobilities
are an order of magnitude smaller than corresponding mobilities for
graphene samples on similar substrates.  We also argue that the
minimum conductivity, plateau width and
threshold voltage have different scaling in graphene and bilayers and
have different dependence on the substrate dielectric constant.  These
differences could potentially be useful when designing a particular
device application.  Moreover, a systematic study of the differences
in transport properties between single layer graphene and bilayer
graphene could verify our claim that the underlying mechanism for most
of the observed transport properties on current graphene (and bilayer)
samples is dominated by charged impurities.

In addition we have argued that both 
graphene and bilayers have weak temperature dependence making
these materials have among the highest room temperature mobility for
any field effect device.  One important consequence for technology is
that since mobility is limited by charged impurities, better samples
can be made either by removing charged impurities in the SiO$_2$
substrate or by using different substrate (e.g. vacuum for the case of
suspended graphene and bilayers).  Attaining high mobility is also
necessary to access new physics such as the fractional quantum Hall
effect or the ``universal'' minimum conductivity, both of which we
believe is not seen in current experiments due to the large number of
charged impurities.

In one sense, bilayer graphene can be thought of as a new material
that shares some properties with regular 2DEGs (e.g. quadratic
Hamiltonian) and some properties with graphene (e.g. chiral
Hamiltonian).  However, theoretically, bilayer graphene is a far more
interesting, where as discussed earlier, modestly increasing the back
gate voltage from that of current experiments should induce a
sufficiently large carrier density to see several interesting effects.
For example, a ``super-linear'' conductivity is predicted both by
high-density corrections in the Thomas-Fermi approximation (within the
quadratic Hamiltonian) as well as by the theoretically expected
crossover between the quadratic and linear Hamiltonians.  At even
higher (and perhaps realistic) densities, one would expect a strong
increase conductivity as one populated higher bands.  The theoretical
framework for understanding these high-density effects (including the
role of chirality) and their observation remains an exciting avenue 
for future theoretical
and experimental research.  In summary, we have proposed a simple
theory for bilayer graphene transport including the effects of
screened Coulomb impurities.  The result of our self-consistent
Drude-Boltzmann semi-classical diffusive transport
theory~\cite{kn:adam2007a} is in good agreement with the recent
experiments of Ref.~\onlinecite{kn:morozov2007} and we make several
predictions that can be tested in future experiments.

We thank Andre Geim for sharing with us his unpublished
data (Ref.~\onlinecite{kn:morozov2007}) and for a careful reading of
our manuscript.  This work is supported by U.S. ONR.

\vspace{0.2in}


\end{document}